\documentclass{iopjournal}
\usepackage{multirow}
\usepackage{amsmath,amsthm,mathtools}
\usepackage{color}
\usepackage{soul}
\usepackage{enumitem}
\usepackage{physics}
\usepackage{dsfont}
\usepackage{footmisc}
\usepackage{cite}
\usepackage{url}
\usepackage{textcomp}
\usepackage{rotating}
\usepackage{lmodern}
\usepackage[dvipsnames]{xcolor}
\usepackage{array}
\usepackage[normalem]{ulem}
\usepackage{tablefootnote}

\definecolor{emerald}{rgb}{0.31, 0.78, 0.47}
\definecolor{blue(ncs)}{rgb}{0.0, 0.53, 0.74}

\hypersetup{
	colorlinks=true,
	linkcolor=blue(ncs),
	filecolor=blue,
	citecolor=emerald,      
	urlcolor =blue(ncs),
}

\definecolor{lightgray}{rgb}{0.9,0.9,0.9}	    
\definecolor{green}{rgb}{0,0.5,0}
\definecolor{red}{rgb}{1,0,0}
\definecolor{blue}{rgb}{0,0,0.5}

\long\def\symbolfootnote[#1]#2{\begingroup%
	\def\thefootnote{\fnsymbol{footnote}}\footnotetext[#1]{#2}\footnotemark[#1]\endgroup}

\usepackage{chngcntr}
\usepackage{graphicx}% Include figure files
\usepackage{dcolumn}% Align table columns on decimal point
\usepackage{bm}

\usepackage{tikz}
\usepackage{tikz-3dplot}
\usetikzlibrary{shapes,arrows, calc, arrows.meta}
\tikzstyle{decision} = [diamond, draw, fill=blue!20, 
text width=4.5em, text badly centered, node distance=3cm, inner sep=0pt]
\tikzstyle{block} = [rectangle, draw, fill=blue!20, 
text width=5em, text centered, rounded corners, minimum height=4em]
\tikzstyle{line} = [draw, -latex']
\tikzstyle{cloud} = [draw, ellipse,fill=red!20, node distance=3cm,
minimum height=2em]

\usepackage{subcaption}     % For sub-figures
\newcommand{\drawB}[2]{
	\begin{scope}[shift={(#1)}]
		\coordinate (origin) at (0,0);  
		
		\draw (origin) -- (1, 0);
		\draw (1, 0) -- (1, 1);
		\draw (origin) -- (0, 1);
		\draw (0, 1) -- (1, 1);
		\draw[thick] (0, 0.5) -- (-0.2, 0.5);
		\draw[thick] (1, 0.5) -- (1.2, 0.5);
		\draw[thick] (0.5, 1) -- (0.5, 1.2);
		\draw[thick] (0.5, 0) -- (0.5, -0.2);

		\node[scale = 0.75] at (0.5, 0.5) {$#2$};
		
	\end{scope}
}

\newcommand{\drawBone}[5]{
	\begin{scope}[shift={(#1)}]
		
		\coordinate (origin) at (0,0);
		
		\node[scale = 0.8] at ($(origin) + (-0.9, 0.5, 0)$) {$#2$};

		\draw[thick, #3, -Stealth, #5] (0.5, 0) -- (1, 0.5);
		\node[scale = 0.6] at (0.4, 0.35) {$#4$};
		
		\draw (origin) -- (1, 0);
		\draw (1, 0) -- (1, 1);
		\draw (origin) -- (0, 1);
		\draw (0, 1) -- (1, 1);

	\end{scope}
	
}

\newcommand{\drawBtwo}[7]{
	\begin{scope}[shift={(#1)}]
		
		\coordinate (origin) at (0,0);
		
		\node[scale = 0.8] at ($(origin) + (-1.3, 0.5, 0)$) {$#2$};
		
		\draw[thick, #3, -Stealth, #5] (0, 0.5) -- (0.5, 1);
		\draw[thick, #3, -Stealth, #6] (0.5, 0) -- (1, 0.5);
		\node[scale = 0.6] at (0.7, 0.75) {$#4$};
		\node[scale = 0.6] at (0.25, 0.18) {$#7$};
		
		\draw (origin) -- (1, 0);
		\draw (1, 0) -- (1, 1);
		\draw (origin) -- (0, 1);
		\draw (0, 1) -- (1, 1);

	\end{scope}
	
}

\newcommand{\drawBthree}[7]{
	\begin{scope}[shift={(#1)}]
		
		\coordinate (origin) at (0,0);
		
		\node[scale = 0.8] at ($(origin) + (-1.3, 0.5, 0)$) {$#2$};
		
		\draw[thick, #3, -Stealth, #5] (0.5, 1) -- (1, 0.5);
		\draw[thick, #3, -Stealth, #6] (0, 0.5) -- (0.5, 0);
		\node[scale = 0.6] at (0.4, 0.75) {$#4$};
		\node[scale = 0.6] at (0.7, 0.18) {$#7$};
		
		\draw (origin) -- (1, 0);
		\draw (1, 0) -- (1, 1);
		\draw (origin) -- (0, 1);
		\draw (0, 1) -- (1, 1);

	\end{scope}
	
}

\newcommand{\drawBfour}[5]{
	\begin{scope}[shift={(#1)}]
		
		\coordinate (origin) at (0,0);
		
		\node[scale = 0.8] at ($(origin) + (-0.9, 0.5, 0)$) {$#2$};

		\draw[thick, #3, -Stealth, #5] (0, 0.5) -- (0.5, 0);
		\node[scale = 0.6] at (0.5, 0.45) {$#4$};
		
		\draw (origin) -- (1, 0);
		\draw (1, 0) -- (1, 1);
		\draw (origin) -- (0, 1);
		\draw (0, 1) -- (1, 1);

	\end{scope}
	
}

\newcommand{\drawBfive}[2]{
	\begin{scope}[shift={(#1)}]
		
		\coordinate (origin) at (0,0);
		
		\node[scale = 0.8] at ($(origin) + (-0.9, 0.5, 0)$) {$#2$};

		\draw (origin) -- (1, 0);
		\draw (1, 0) -- (1, 1);
		\draw (origin) -- (0, 1);
		\draw (0, 1) -- (1, 1);

	\end{scope}
	
}

\newcommand{\drawNoArrowBC}[2]{
	\begin{scope}[shift={(#1)}]
		
		\coordinate (origin) at #1;
		\coordinate (endpoint) at #2;
		\fill[black] (endpoint) circle (3pt);
		\draw[thick] (origin) -- (endpoint);

	\end{scope}
}
\usepackage[nameinlink,noabbrev]{cleveref} % For clever 
\crefformat{equation}{#2Eq.~\textup{(\textcolor{blue}{#1})}#3}
\Crefformat{equation}{#2Equation~\textup{(\textcolor{blue}{#1})}#3}

\crefformat{figure}{#2Fig.~\textup{(\textcolor{blue}{#1})}#3}
\Crefformat{figure}{#2Figure~\textup{(\textcolor{blue}{#1})}#3}

\crefformat{table}{#2Table.~\textup{\textcolor{blue}{#1}}#3}
\Crefformat{table}{#2Table~\textup{\textcolor{blue}{#1}}#3}

\begin{document}

\articletype{Paper} %	 e.g. Paper, Letter, Topical Review...

\title{Entanglement Scaling of Motzkin Chains under Spiral Hamiltonian-path Embeddings}

\author{Zhao Zhang$^{1,*}$\orcid{0000-0002-9425-732X},  and Olai B. Mykland$^{1,2,3}$}

\affil{$^1$Department of Physics, University of Oslo, P.O. Box 1048 Blindern, N-0316 Oslo, Norway}

\affil{$^2$Department of Computer Science, University of Copenhagen, Copenhagen, Denmark}

\affil{$^3$Department of Physics, Norwegian	University of Science and Technology, NO-7491, Trondheim, Norway}

\affil{$^*$Author to whom any correspondence should be addressed.}

\email{zhao.zhang@fys.uio.no}


\begin{abstract}
We investigate how a two-dimensional spatial embedding changes the entanglement observed across geometrically local bipartitions of a one-dimensional quantum spin chain. A $q$-deformed colored Motzkin chain is placed along a spiral-inspired Hamiltonian path that visits $N(L)=\Theta(L^2)$ sites of a square-lattice region while retaining a path interaction graph. Straight ambient-space cuts then correspond to unions of $\Theta(L)$ disconnected intervals in the chain ordering. We derive the scaling of a height-variation proxy for the leading color-pairing contribution and predict ambient-space entropy scaling proportional to $L^2$, $L$, and $\mathcal{O}(1)$ for $q>1$, $q=1$, and $q<1$, respectively. We evaluate this proxy numerically using the matrix-product-state representation and verify its predicted finite-size scaling. The embedded and original Hamiltonians have exactly identical spectra and spectral gaps at fixed chain length. We also describe a sparse tensor-network representation and propose a two-chain junction extension. These results isolate the roles of interaction connectivity, spatial embedding, and subsystem geometry in entanglement scaling.
\end{abstract}

\section{Introduction}
\label{sec:Intro}
%%%
%%
%
Entanglement entropy (EE) quantifies how strongly a quantum state deviates from being classical. In this sense, the ground state (GS) of a gapped, locally interacting quantum many-body system is relatively unentangled compared with a typical eigenstate in the full Hilbert space \cite{DonPage}: The EE of a subsystem scales with the size of its boundary rather than its volume. This is the area law of EE \cite{EisertAreaLaw}, which is expected to hold for any spatial dimension. The intuition is that a spectral gap between the GS and first excited state corresponds to massive quasiparticles with exponentially decaying correlations. Hence, only degrees of freedom within a finite range across the boundary contribute significantly to bipartite EE. In one dimension (1D), this intuition has been made rigorous using Lieb–Robinson bounds \cite{Hastings_2006}. In two dimensions (2D), progress has recently been made from certain classes of frustration-free Hamiltonians \cite{GossetSubvol,Anshu_2022}, a family that includes the model proposed in this work, to a general finite-range Hamiltonian \cite{OAI:A-two-dimensional-area-law-from-a-global-spectral-gap-September-24-2026}.

However, the area law does not constrain how entangled a GS can become when the spectral gap closes, as in quantum critical systems. At such critical points, EE typically acquires an additional logarithmic enhancement beyond the area law. This behavior is well understood in (1+1)D using conformal field theory \cite{HOLZHEY1994443,Calabrese_2009} and for gapless free fermions with a Fermi surface in arbitrary dimensions via the Widom conjecture \cite{PhysRevLett.96.100503,PhysRevLett.96.010404}. This enhancement is natural given that the correlation length diverges at criticality, constrained only by the finite size effect. More surprising is that, when the spectral gap closes even more rapidly with system size, for instance exponentially rather than with power law, the GS can exhibit volume scaling of EE. This phenomenon was first demonstrated in an inhomogeneous XX spin chain \cite{Vitagliano_2010,Ramírez_2014,Ramírez_2015}, the GS of which can be approximated as a collection of Bell pairs symmetrically arranged about the center, leading to maximal EE; this model is known as the rainbow chain.

Frustration-free Hamiltonians have enabled closely related constructions of highly entangled GSs with more explicit analytic control \cite{PhysRevLett.109.207202}, including models that are translationally invariant up to boundary effects. For such Hamiltonians, the GS minimizes the energy of each local interaction term individually. Therefore, much of its structure can be deduced by diagonalizing the local Hamiltonian on a small number of sites. In the Motzkin spin chain \cite{PhysRevLett.109.207202}, spin-1 configurations are mapped to discrete random walks, and the GS is expressed as a superposition of Motzkin paths, which begin and end at height zero and never go below the baseline. Bravyi et al.~showed that this GS exhibits EE, correlation functions, and spectral gaps with the same scaling behavior as standard critical systems. Movassagh and Shor extended this construction by adding a color degree of freedom to the local Hilbert space, yielding an EE that grows with subsystem size according to a power law that dominates over the logarithmic scaling EE contributed by the spin degrees of freedom \cite{ShorEntanglement}. These ideas were subsequently adapted to spin-$\frac{1}{2}$ systems with smaller local Hilbert spaces but involving three-body interactions among next-nearest neighbors \cite{PhysRevB.94.155140,FredkinSpinChain}, giving rise to the Fredkin chain. To obtain volume scaling of EE in the GS, these models must be deformed in a frustration-free manner so that the GS weight is biased toward higher paths, which support more Bell pairs across a given cut \cite{ZhaoNovelPT,DeformedFredkinChain,DeformedFredkinChain_explanation}. Tuning the deformation parameter then drives an entanglement phase transition (EPT) from an area law to a volume scaling. In addition to being GSs of local Hamiltonians, such highly entangled states can also be generated using push-down automata \cite{Gopalakrishnan_2025}.

Because an appropriate amount of entanglement is a key resource for quantum computation, and realistic architectures are expected to be implemented on 2D chip layouts rather than 1D wires, it is important to understand, at a fundamental level, how highly entangled GSs can be generated in two dimensions and how spatial embeddings of one-dimensional quantum states affect their entanglement properties.
Here, we study a specific embedding of a one-dimensional frustration-free spin chain into a two-dimensional square-lattice layout, without introducing interactions beyond those along the chain. As we show below, this embedding is predicted to produce distinctive ambient-cut entropy scaling and has an exactly related spectral gap.

Extending the Bell-pair picture from 1D to 2D is, however, highly nontrivial. In 2D, entanglement along different directions competes due to entanglement monogamy. Simple attempts to generalize the rainbow chain have been explored in Refs.~\cite{Ramírez_2015,ZHANG2023169395}, but these either produce volume scaling of EE only along a single lattice direction, or do so on structures of Hausdorff dimension one embedded in 2D. A similar trade-off between entanglement in orthogonal directions appears in colored extensions of the toric code \cite{balasubramanian20232dhamiltoniansexoticbipartite,Zhang2024bicolorloopmodels}, where enhanced boundary entanglement leads to anomalous topological EE.

A further challenge is defining a consistent notion of `height' in 2D, since height differences between two points can depend on the path taken. A systematic resolution of both issues was achieved by coupling two perpendicular arrays of 1D Motzkin or Fredkin chains with appropriate vertex or tiling rules, which enforce an emergent Coulomb-gauge-like constraint and yield an unambiguous height function \cite{ZhaoSixNineteenVertex,Zhang2024quantumlozenge}. The resulting models remain frustration free, and the couplings do not disrupt the entangling interactions along each chain. In contrast to anisotropic coupled-wire constructions of topological superconductors \cite{PhysRevB.94.165142}, where inter-wire couplings open a gap, these systems stay gapless. Their GSs can be viewed as superpositions of discrete random membranes above a hard wall at zero height, with amplitudes weighted by a deformation parameter raised to the volume beneath the membrane. The same EPT persists, but reduced height fluctuations in higher dimensions lead to a logarithmic violation of the area law at criticality, rather than the power-law violation found in the 1D models.

While the six-/19-vertex and lozenge-tiling realizations are natural 2D analogues of the Motzkin and Fredkin chains, their local interactions act on four or six sites, respectively, versus two or three in 1D. Although the framework generalizes in principle to higher dimensions, the required interaction range becomes increasingly difficult to realize. A sequentially generable 2D state with high entanglement along one direction has also been constructed, but its local terms act on eight sites \cite{zhang2025sequentialgenerationtwodimensionalsuperarealaw}. Another proposal is a `loop-soup' construction \cite{balasubramanian20232dhamiltoniansexoticbipartite}, where the GS is a superposition of random loop configurations, each loop hosting a periodic Motzkin chain \cite{PRONKO2025116963}. However, the randomness of the loops obscures the scaling of EE for generic bipartitions, and a naive periodic extension of the Motzkin Hamiltonian leads to an exponentially degenerate GS \cite{PRONKO2025116963}, making it trivial to obtain exotic EE scaling in conjunction with high degeneracy.

In this work, we investigate how embedding a one-dimensional frustration-free spin chain into a two-dimensional square-lattice layout affects the EE between subsystems and the scaling of the spectral gap. The key idea is to embed a single Hamiltonian path that visits each site of a 2D lattice exactly once, without self-intersections. Such paths have appeared before in 2D generalizations of the Jordan–Wigner transformation \cite{PhysRevLett.63.322,PhysRevLett.86.1082}, where a zigzag ordering is conventional. However, a zigzag path would concentrate strong entanglement along only one lattice direction. Instead, a Fermat-spiral-inspired Hamiltonian path is motivated by the Bell-pair structure of the highly entangled ground state: Sites paired symmetrically about the center of the one-dimensional chain are mapped to lattice sites related by point reflection about the center of the square. This arranges the chain's long-range entanglement in a centrally symmetric pattern within the 2D layout. This construction can be viewed as a strongly anisotropic nearest-neighbor Hamiltonian whose active bonds follow the spiral path, rather than a fully coupled 2D lattice Hamiltonian

Because the locations of the entangling degrees of freedom are fixed along a single path, the EE scaling becomes analytically tractable using known asymptotic results for the EE of Motzkin-chain segments whose lengths form arithmetic progressions. As in previous coupled-chain constructions, the proxy predicts volume-scaling bipartite EE for any cut through the center. In contrast to those models, however, the present construction obeys an area law at the critical point and exhibits subsystem-size-independent EE in the weakly entangled phase. Similar relations between the bipartite EE scaling of 1D systems and twisted 2D spirals with alternating segments may apply for other highly entangled chains such as the rainbow chain, and a conjectured leading-exponent conversion is formulated.

The scaling of EE is tightly linked to the structure of tensor network (TN) representations of quantum states. A hierarchical TN for the Motzkin GS with logarithmic area-law violation was proposed in Ref.~\cite{Alexander2021exactholographic} and was recently shown to be equivalent to a generalized multiscale entanglement renormalization ansatz (MERA) \cite{mykland2025reconcilingtranslationalinvariancehierarchy}. This equivalence was established by factorizing a rank-5 tensor, i.e.~the minimal unit cell of the TN, into products of rank-3 tensors and recombining them. Here, inspired by the recently introduced holographic TN for the double-scaled SYK model \cite{10.21468/SciPostPhys.19.4.083}, we perform an analogous tensor reshaping for the highly entangled phase of the Fredkin chain. By enforcing conservation of arrow flow, we further reduce the tensor support to a small set of configurations that alternate between even and odd coordinates in the 2D bulk of the holographic TN. We then adapt this optimized TN to describe the GS of the embedded 1D chain 
and compare its global structure with the genuine 2D TN previously obtained for the six-vertex-coupled Fredkin chains \cite{mykland2025highlyentangled2dground}.

Although the construction is based on a single Hamiltonian path and therefore remains 1D in its interaction connectivity, the spiral embedding admits a natural junction-like generalization. In this regard, it is useful to recall that Fermat’s spiral can be viewed as a special case of the spiral meshes appearing in disc phyllotaxis. Our approach naturally generalizes to a mesh of spirals by introducing two orthogonal Motzkin/Fredkin chains, each winding as a Fermat spiral and intersecting at the lattice center. This structure connects to the well-studied physics of quantum wire junctions with Hausdorff dimension one \cite{wirejunction,CRAMPE2013526,Juhász_2018,fddq-8lzl}, but now realized within a 2D lattice. We propose a junction-like extension that may generate long-range entanglement between distant spins on different branches of the junction and isotropic volume-scaling EE for bipartitions of the embedded lattice.

The remainder of this article is organized as follows. In Sec.~\ref{sec:model} we briefly review the key features of the Motzkin chain, including its GS and parent Hamiltonian. Sec.~\ref{sec:Fermat} introduces the Fermat-spiral layout of a Hamiltonian path on the square lattice, on which the Motzkin Hamiltonian is embedded. In Sec.~\ref{sec:EE} we analyze the height-variation proxy and formulate the corresponding entropy-scaling prediction by mapping 2D bipartitions to disjoint segments along the 1D chain, and compare the resulting EE and spectral-gap behavior across phases and critical points to those of earlier coupled-chain constructions. In Sec.~\ref{sec:TN} we introduce a new TN representation of the highly entangled 1D chains and compare its 2D `twisted' version to a fully 2D TN with closely related global geometry. The Fermat-spiral construction is further generalized to a centrally symmetric quantum wire junction in Sec.~\ref{sec:junction}. We conclude in Sec.~\ref{sec:Concl} with perspectives on higher-dimensional generalizations and on EE scaling laws for gapless GSs.

\section{The Motzkin Hamiltonian and ground state}
\label{sec:model}

\begin{figure*}[ht]
	\centering
	\includegraphics[width=0.6\linewidth]{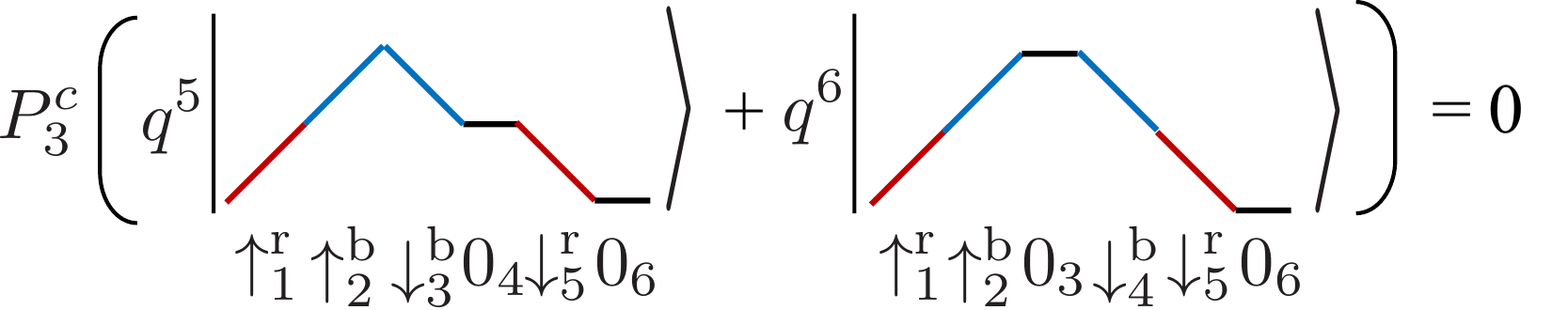}
	\caption{A demonstration of how the projection operator acting on site 3 and 4 of any color annihilates the specific superposition of colored spin configurations weighted by $q$ to the power of the area under the corresponding Motzkin path, which appears in the GS.}
	\label{fig:GS}
\end{figure*}
The $q$-deformed $s$-colored Motzkin chain~\cite{ZhaoNovelPT} has a $(2s+1)$-dimensional local Hilbert space, with spin up and down, both in any of the $s$ colors, and an uncolored spin zero. The Hamiltonian is expressed in terms of projectors 
\begin{equation}
	P_{j}^c=|L^c\rangle_{j,j+1}\langle L^c|_{j,j+1}+|R^c\rangle_{j,j+1}\langle R^c|_{j,j+1}+|F^c\rangle_{j,j+1}\langle F^c|_{j,j+1},
\end{equation}with
\begin{equation}
	\begin{split}
		|L^c\rangle_{j,j+1} =& \frac{1}{\sqrt{1+q^2}}\left(|\uparrow^{c}_{j}0_{j+1}\rangle - q|0_{j}\uparrow^{c}_{j+1}\rangle\right),\\
		|R^c\rangle_{j,j+1} =& \frac{1}{\sqrt{1+q^2}}\left(|0_{j}\downarrow^{c}_{j+1}\rangle - q|\downarrow^{c}_{j}0_{j+1}\rangle\right),\\
		|F^c\rangle_{j,j+1} =& \frac{1}{\sqrt{1+q^2}}\left(|\uparrow^{c}_{j}\downarrow^{c}_{j+1}\rangle- q|0_{j}0_{j+1}\rangle \right),
	\end{split}
	\label{eq:projectors}
\end{equation}where $c=1,2,\cdots, s$ is the color index, and $q>0$ is the deformation parameter. Each of the projection operators has a local lowest energy eigenstate corresponding to the direction orthogonal to the above three vectors, as demonstrated in \cref{fig:GS}. So the GS of a Hamiltonian consisting of such projectors acting on all neighboring pairs of sites in the chain is a weighted superposition of random walks if the up, down and 0 spins are respectively mapped to up, down and flat moves. In addition to the color matching encoded in $|F^c\rangle_{j,j+1}$ between up and down moves when they are pair created from spin zeros, an energy penalty needs to be introduced by the projector $C_{j}^{c_1,c_2}=|\uparrow^{c_1}_{j}\downarrow^{c_2}_{j+1}\rangle\langle \uparrow^{c_1}_{j}\downarrow^{c_2}_{j+1}|$ for $c_1\ne c_2$ in order to enforce color matching between nearest up-down pairs in the GS configurations. The remaining GS degeneracy can be lifted by fixing the boundary conditions with the Hamiltonian $H_\partial=\sum_{c=1}^s|\downarrow^{c_1}\rangle\langle\downarrow^{c_1}|_{1}+|\uparrow^{c_1}\rangle\langle\uparrow^{c_1}|_{2N}$, for a chain of length $2N$. In summary, the Motzkin Hamiltonian is given by
\begin{equation}
	H= H_\partial+\sum_{j=1}^{2N-1}\left(\sum_{c=1}^sP_{j}^c+\sum_{c_1\ne c_2}C_{j}^{c_1,c_2}\right).
\end{equation}

The unique GS of this Hamiltonian is a superposition of all spin configurations where there are always at least as many up spins as down spins to the left at any point along the chain, and the colors of the nearest up-down spin pairs are either both red or both blue, in a nested fashion. If up (resp. down) spins are mapped to $45^\circ$ upward (resp. downward) moves with unit horizontal lengths, and the area enclosed between a Motzkin path $m$ and the horizontal axis at height zero is denoted $A(m)$, then up to an overall normalization factor, the GS can be expressed as
\begin{equation}
	|\mathrm{GS}\rangle=\sum_{m\in M_{2N}}q^{A(m)}|m\rangle,
\end{equation}where $M_{2N}$ denotes the set of bicolor Motzkin paths of length $2N$ with color matching. 

The most striking property of such a GS is that as the deformation parameter $q$ is tuned, an EPT happens in the thermodynamic limit of $N\to\infty$. For $q<1$, the superposition is dominated by lower Motzkin paths. As the major contribution to the EE comes from the matching of colors in nearest up-down spin pairs, the bipartite EE across the center of the chain is bounded by a constant regardless of the subsystem size. When $q>1$, the height of the Motzkin paths at the center on average scales linearly with the system size. So most of the spins in a half system are entangled with those in the other, and the EE therefore has a volume scaling. At the critical point $q=1$, there are two factors that are competing one another. The first is the hard-wall constraint from zero height, pushing the average Motzkin paths away from it. The other is the entropy, or number of microscopic configurations for a coarse-grained height profile, which obviously grows as the path gets flatter and closer to the baseline. So the exact scaling behavior in this case is determined by the balance of these two factors, which turns out to be similar to that of a 1D random walk without the Motzkin property \cite{Zhang2024quantumlozenge}. More detailed accounts and rigorous proofs regarding the Motzkin chain can be found in Refs.~\cite{PhysRevLett.109.207202,ShorEntanglement,ZhaoNovelPT}, see also its half-integer spin counterpart, the Fredkin chain \cite{PhysRevB.94.155140,FredkinSpinChain,DeformedFredkinChain,DeformedFredkinChain_explanation}. Here, we simply summarize the common scaling behavior of their bipartite EE across the center and spectral gaps in \cref{tab:table0}, which will be used in the next sections to obtain the scaling behavior of the spiral model.

\begin{table}[ht]
	\centering
	\caption{\label{tab:table0}%
		Summary of previous results on the bipartite EE and spectral gap scaling behavior for the 1D colored and colorless Motzkin and Fredkin chains.
	}\renewcommand{\arraystretch}{1.5}
	\begin{tabular}{|c |c|c|c |}\cline{1-4}
		\textrm{Model}&
		\textrm{Phase}& \textrm{Bipartite EE}&
		\textrm{Spectral gap} \\ \cline{1-4}
		& $q>1$ & $N$ \cite{ZhaoNovelPT,DeformedFredkinChain,DeformedFredkinChain_explanation} & $e^{-\beta N^2}$~\cite{Levine_2017,DeformedFredkinChain_explanation}   \\ \cline{2-4}
		$s\ge 2$ & $q=1$  & $\sqrt{N}$ \cite{ShorEntanglement,FredkinSpinChain}  & $N^{-\delta}$~\cite{ShorEntanglement} \\ \cline{2-4}
		& $q<1$ & $\mathcal{O}(1)$\cite{ZhaoNovelPT,DeformedFredkinChain,DeformedFredkinChain_explanation} & $\mathcal{O}(1)$ \cite{Andrei:2022yym} \\      \cline{1-4}
		& $q>1$ & $\mathcal{O}(1)$\cite{ZhaoNovelPT,DeformedFredkinChain,DeformedFredkinChain_explanation} & $e^{-\zeta N}$ \cite{Levine_2017,DeformedFredkinChain_explanation}\\ \cline{2-4}
		$s=1$ & $q=1$ & $\log N$ \cite{PhysRevLett.109.207202,FredkinSpinChain} &  $N^{-\theta}$~\cite{PhysRevLett.109.207202} \\ \cline{2-4}
		& $q<1$ & $\mathcal{O}(1)$\cite{ZhaoNovelPT,DeformedFredkinChain,DeformedFredkinChain_explanation} & $\mathcal{O}(1)$ ~\cite{Andrei:2022yym}\\
		\cline{1-4}
	\end{tabular}
\end{table}

\section{Fermat's spiral as a Hamiltonian path on the square lattice}\label{sec:Fermat}

\begin{figure*}[ht]
	\centering
	\includegraphics[width=\linewidth]{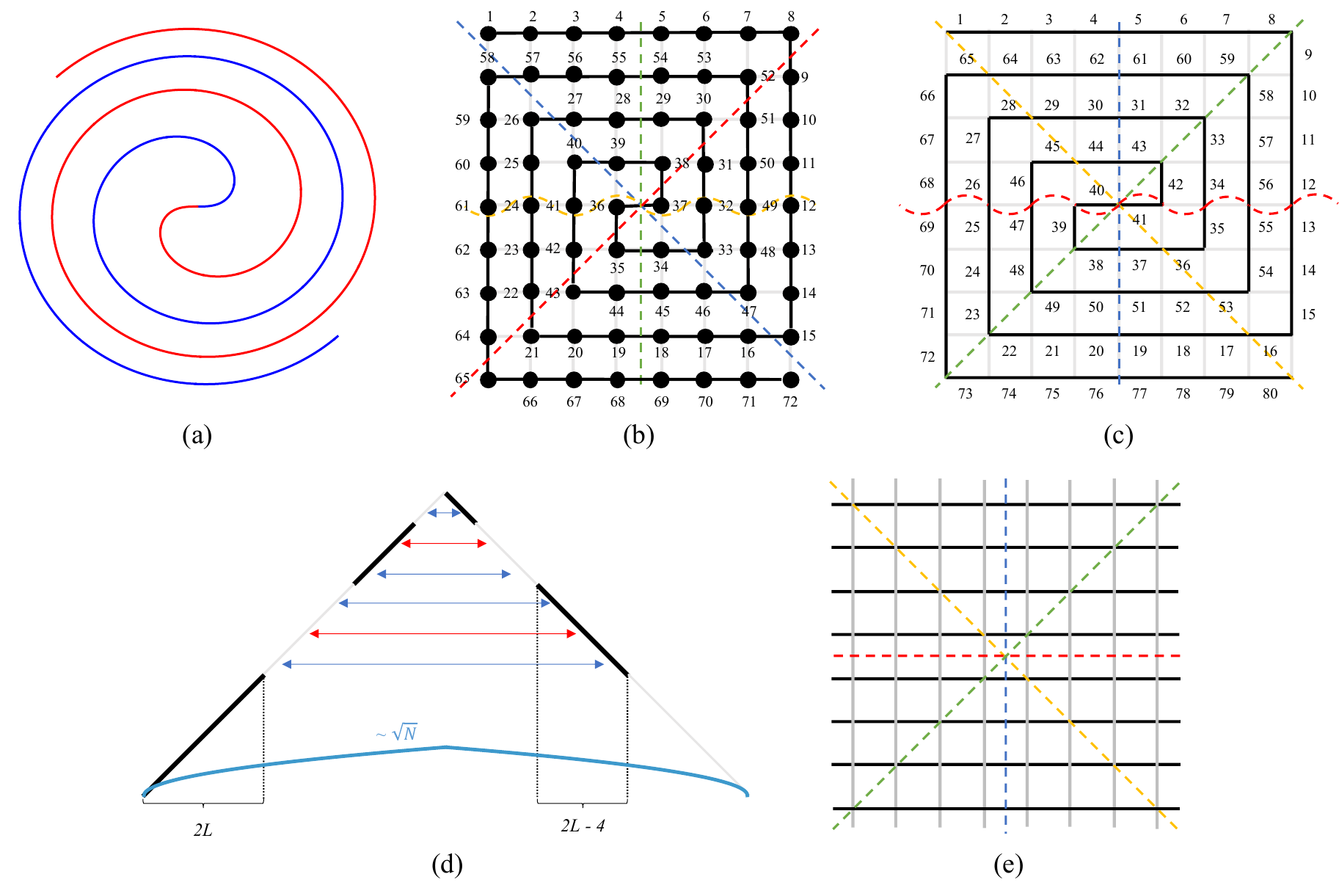}
	\caption{(a) Fermat's spiral plotted with the positive and negative branches in different colors. (b) An $L\times (L+1)$ square lattice with $2N=L(L+1)$ spins on the vertices, their interaction is turned on along a spiral-like Hamiltonian path. (c) An $L\times L$ square lattice where the $2N=L(L+2)$ spins live on the edges instead. Four symmetric bipartition cuts are shown in different colors. (d) The two subsystems resulting from the yellow cut in (c). The northeast subsystem is represented with the thickened black lines, whereas the southwest is marked by thin gray lines. This is asymptotically equivalent to the yellow bipartition in (b), separating the upper half plane from the lower. The configuration depicted corresponds to the probabilistically dominant Motzkin path in the GS superposition for the $q>1$ phase, to illustrate the entanglement across subsystems. Typical height configuration at the critical point is depicted in blue. (e) The uncoupled arrays of orthogonal chains with different EE and gap scaling for comparison in \cref{tab:table1} and \cref{tab:table2}.}
	\label{fig:snake}
\end{figure*}

The Fermat spiral in polar coordinates is given by the equation $r=\pm \sqrt{\varphi}$ for $\varphi>0$. The plus and minus signs give the two branches of the spiral meeting at the origin, as shown in \cref{fig:snake} (a). The Fermat curve has been widely applied to model plant growth and the shape of spiral galaxies, as well as in the design of variable capacitors, concentrated solar panels and cyclotrons. In graph theory, a Hamiltonian path is a path that visits each vertex of the graph exactly once. On regular lattices such as the square lattice, it is a topic well studied in statistical mechanics and quantum many-body physics as fully-packed loop models \cite{Zhang_2023}. In \cref{fig:snake} (b) and (c), we depict two such paths that mimic Fermat's spiral. The difference between them is that in \cref{fig:snake} (b) we define the spin degrees of freedom on the vertices of the lattice, while in (c) they are defined on the edges. In either case, we impose the Motzkin interaction between neighboring spins on the Hamiltonian path of the lattice, and completely turns of the interaction with the other two neighboring sites that are not adjacent on the path. Depending on the location of the spin on the square lattice, the interaction can be either between east-west neighbors for the upper and lower quadrants, or between the north-south neighbors in the left and right quadrants. Near the two diagonal lines, where the Hamiltonian path turns, the interaction is between neighbors around its corner. The two layouts of the spins and neighboring interactions are asymptotically equivalent in terms of the scaling behavior of bipartite EE. However, the latter has a more regular distribution of segments along the Motzkin chain, so for convenience we will work with this lattice in the following.

To study the isotropic EE, four symmetric bipartition cuts are shown in \cref{fig:snake} (b) and (c). They separate the spins into two subsystems consisting of increasingly larger segments of size $4n-a$, for $n=1,2,\cdots,L/2+1$, with $a=0,1,2,3$ corresponding respectively to the yellow, blue, green and red cut. Starting from the center, segments within one subsystem in the 2D lattice on a given side of the cut alternate between the first and second halves of the 1D chain when it is stretched straight. Since the scaling behavior of bipartite EE for any of the cuts will not differ qualitatively, from now on we focus on the yellow cut along the main diagonal with $a=0$, which results in a bipartition shown in \cref{fig:snake} (d).

\section{Scaling of entanglement entropy and spectral gap}
\label{sec:EE}

As analyzed in Ref.~\cite{ZhaoNovelPT}, the EE of the Motzkin GS decomposes into a leading contribution from the color degree of freedom, and a subleading contribution from the spin sector. Only up-down spin pairs belonging to different subsystems account for the former. In the Motzkin path picture, this means that only the net height change of a subsystem, or the surplus of up spins matters. For the 1D system studied there, the leading contribution to the EE is therefore proportional to the height at the center of the Motzkin path, giving the three distinct scaling behaviors in the two different phases and the critical point. For the critical point at $q=1$, the EE of a single connected segment in the middle of the chain has also been obtained from combinatorial analyses for the colorless \cite{10.1063/1.4977829} and colored Motzkin chains \cite{Menon:2024vic}. Here, the situation seems a bit more complicated, as segments belonging to two subsystems alternate along the chain. A rigorous estimation would rely on the enumeration of Motzkin-like paths that stay within the upper half plane but starts and ends at nonzero height. In \cref{sec:reflection}, we point out the challenge of a rigorous estimation from the combinatorial and path-integral perspective.

The decomposition of EE into a color contribution and a height fluctuation contribution can be seen from the Schmidt decomposition of the GS
\begin{equation}
	|\mathrm{GS}\rangle=\sum_{c,\bm{\delta h}}\sqrt{p_{\bm{\delta h},c}}|A_{\bm{\delta h},c}\rangle\otimes|B_{\bm{\delta h},c}\rangle.
\end{equation}
The Schmidt coefficient $p_{\bm{\delta h},c}$ depends on the total height changes $\bm{\delta h}= [\delta h^{1}, \delta h^{2}, ..., ]$, where each entry refers to the height change of a segment resulting from the cut across the 2D lattice, and the coloring $c$ of the surplus up- or down-spins that are paired between the two subsystems $A$ and $B$. Since the number of such pairs across subsystems depends on the height changes, the joint probability $p_{\bm{\delta h},c}$ is the marginal probability $p_{\bm{\delta h}}=\sum_{c}p_{\bm{\delta h},c}$ times the conditional probability $p_{c|\bm{\delta h}}=p_{\bm{\delta h},c}/p_{\bm{\delta h}}$. The resulting entropy can therefore be written as
\begin{equation}
	\begin{split}
		S(\{p_{\bm{\delta h},c}\})=&-\sum_{c,\bm{\delta h}}p_{\bm{\delta h},c}\log p_{\bm{\delta h},c}\\
		=&-\sum_{c,\bm{\delta h}}p_{\bm{\delta h}}p_{c|\bm{\delta h}}\left(\log p_{\bm{\delta h}}+\log p_{c|\bm{\delta h}}\right)\\
		=&-\sum_{\bm{\delta h}}p_{\bm{\delta h}}\sum_c p_{c|\bm{\delta h}}\log p_{c|\bm{\delta h}}-\sum_{\bm{\delta h}}p_{\bm{\delta h}}\log p_{\bm{\delta h}}\sum_{c}p_{c|\bm{\delta h}}\\
		\equiv&\sum_{\bm{\delta h}}p_{\bm{\delta h}} S_c(\bm{\delta h})-\sum_{\bm{\delta h}}p_{\bm{\delta h}}\log p_{\bm{\delta h}}\\
		=&\langle S_c\rangle+S_h.
	\end{split}
\end{equation}Assuming that the spins in each segment are always paired with those in the other subsystem, instead of other segments in the same subsystem, with the opposite sign of total spin, the conditional probability is simply $p_{c|\bm{\delta h}}=\prod_s 2^{-\delta h^s}$, where the superscript $s$ labels the segments in subsystem $A$. This follows from the fact that the two colors are equally likely. This is indeed the paths with the largest $p_{\bm{\delta h}}$ in the scaling limit, as the height function is monotonic in the two halves of the chain, as shown in \cref{fig:snake} (c). Motivated by the monotone mean-height profile in each half of the chain, we adopt the approximation that the number of matched pairs connecting disconnected intervals within the same subsystem is subleading compared with the total interval height variation. Conditional on this assumption, we may estimate the leading contribution by 
\begin{equation}
	\langle S_c(\bm{\delta h})\rangle =-\sum_c \langle p_{c|\bm{\delta h}}\log p_{c|\bm{\delta h}}\rangle =\log 2 \sum_s|\delta h^s|.
\end{equation}This is in proportion to the total height change in the subsystem.

Instead of evaluating the path integral \cref{eq:pathintegral} by brute force, we will use the known results on the asymptotic scaling of the height function. For $q=1$, the height at distance $x$ away from the endpoint $h_{k} = \sum^{k}_{i = 1}S^{z}_{i}$ on average scales as $\langle h_k\rangle \sim \sqrt{k}$ in the continuous limit \cite{ZhaoNovelPT,DeformedFredkinChain}, which is due to the repellent hard wall at zero height. For the particular bipartition in question, we can further use the symmetry about the midpoint $\langle h_k\rangle=\langle h_{L-k}\rangle$ to write \cref{eq:pathintegral} as
\begin{equation}
	S\sim \sum_{k=1}^{\frac{L}{2}}(-1)^{k}\langle h_{k-1}\rangle+\sum_{k=\frac{L}{2}}^{L-1}(-1)^{k}\langle h_{L-k}\rangle=\langle h_{\frac{L}{2}}\rangle=\sqrt{\sum_{j=0}^{\frac{L}{2}-1}(2L-4j)}=\sqrt{L^2+2L},\label{eq:Shrelation}
\end{equation}
which scales linearly with $L$ asymptotically. In fact, this is nothing but the average height at the middle of the chain, meaning that the bipartite EE of the 2D lattice of linear size $L$ scales in the same way as the EE of half chain with length $L^2$.

For the $q>1$ phase, $\langle h_k\rangle \sim k$, so a completely analogous argument gives the scaling $S\sim L^2$. The $q<1$ phase, on the other hand, has a bounded height everywhere $\langle h_k\rangle \sim \mathcal{O}(1)$, which results in the scaling $S\sim \mathcal{O}(1)$. Notice that this EE is significantly sub-area law, reminiscent of the colorless version of the coupled chain models \cite{Zhang2024quantumlozenge, ZhaoSixNineteenVertex}, and more severe than the topological and logarithmic corrections of the colored loop models studied in Ref.~\cite{Zhang2024bicolorloopmodels}. The above discrete analysis can be much more succinctly summarized in the conversion formula between the bipartite EE $S_1(L)$ of the 1D straight chain of size $2L$ and $S_2(L)$ of the 2D spiral system with $2N\sim L^2$ spins and of linear size $L$ in the continuous limit
\begin{equation}
	S_2(L)\sim S_1(L^2). \label{eq:con1}
\end{equation}Equation \eqref{eq:con1} summarizes the proposed leading-exponent conversion. Its validity requires that color pairs connecting intervals within the same subsystem contribute subleading corrections, an assumption tested numerically below. From this relation, we obtain the new results in \cref{tab:table0}.

\begin{figure*}[t]
	\centering
	% --- Fredkin (a) and (b) ---
	\begin{subfigure}[t]{0.32\linewidth}
		\includegraphics[width=\linewidth]{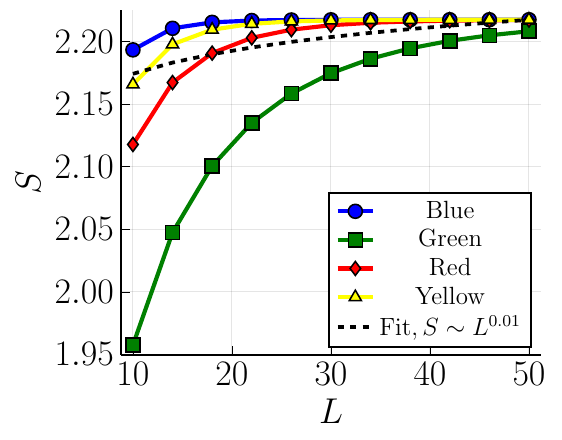}
		\caption{}
	\end{subfigure}
	\hfill
	\begin{subfigure}[t]{0.32\linewidth}
		\includegraphics[width=\linewidth]{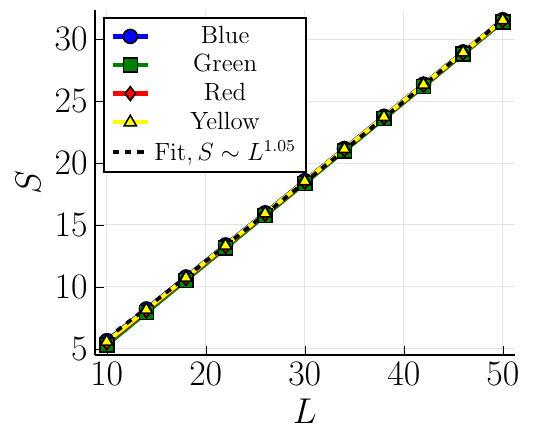}
		\caption{}
	\end{subfigure}
	\hfill
	\begin{subfigure}[t]{0.32\linewidth}
		\includegraphics[width=\linewidth]{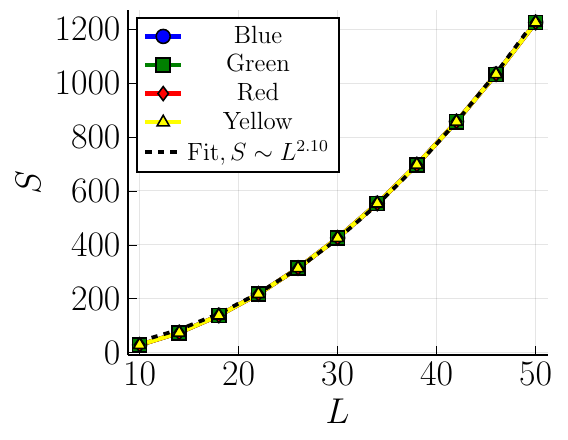}
		\caption{}
	\end{subfigure}
	\caption{Scaling of the bipartite EE $S$ computed from \cref{eq:S_scaling_eq_MPS} and the MPS representation of the GS for various linear system sizes $L$ and for the different colored cuts in \cref{fig:snake} (c). In (a) $q = 0.99$, in (b) $q = 1$ and in (c) $q = 1.01$.}
	\label{fig:S_scaling_MPS}
\end{figure*}

To cross-check the above estimation of the EE scaling in the various phases of the Motzkin GS, we can also use its matrix product state (MPS) representation, found in Ref. \cite{mykland2025reconcilingtranslationalinvariancehierarchy}. 
By summing up the absolute value of the height differences $\Delta_{s}$ across segments $s$ in one of the subsystems defined by one of the cuts in \cref{fig:snake}, we can estimate the EE $S$. That is

\begin{equation}
	S\sim \sum_{s\in{\mathcal{S}_{\text{cut}}}}|\Delta_{s}|=\sum_{s\in{\mathcal{S}_{\text{cut}}}}|\langle h^{s}_{f} - h^{s}_{i}\rangle|,
	\label{eq:S_scaling_eq_MPS}
\end{equation}where $\mathcal{S}_{\text{cut}}$ contains all segments in one of subsystems defined by one of the colored cuts in \cref{fig:snake} (c) and $h^{s}_{i}$ (resp. $h^{s}_{f}$) is the height at the beginning (resp. end) of the segment $s$. For example, if we consider the blue cut in \cref{fig:snake} (c) and the northeast subsystem, $\mathcal{S}_{\text{cut}}$ will consist of segments $[1\dots15]$, $[47\dots57]$, $[27\dots33]$ and $[37, 38, 39]$. In each of the segments, $h^{s}_{i}$ and $h^{s}_{f}$ is the height at the spin denoted by the first and the last number respectively. Note that \cref{eq:S_scaling_eq_MPS} is identical with the first form of \cref{eq:pathintegral}. The expected height values $\langle h\rangle$ are computed from the MPS representation. This gives the plots in \cref{fig:S_scaling_MPS}. The proxy displays exponents consistent with the proposed entropy scaling, but this comparison does not by itself constitute a calculation of the entropy. The results are also independent of the exact bipartition considered in the large $L$ limit.

The subleading contribution to the EE is from the fluctuation in the height of Motzkin paths, which is also the sole source of entanglement for a monochromatic version of the model. A Schmidt decomposition of the reduced density matrix shows that the two subsystems only depend on each other via the total spin within one subsystem, such that the total spin of the whole chain is zero. %Moving from the 1D system to our 2D construction,
Embedding the 1D chain in the 2D lattice, nothing changes except the subsystem is distributed into disconnected segments. So the dimensionality of the reduced density matrix is still the number of possible values of the total spin. However, in the $q>1$ (resp. $q<1$) phase, the probability of the total spin, or height change decays exponentially with the deviation of the area underneath a Motzkin path from the most probable one that is of highest height (resp. completely flat)~\cite{ZhaoNovelPT,DeformedFredkinChain}. So only paths with a height change at finite distance away from the typical one effectively matter, giving a contribution to the EE from height fluctuations that is a bounded constant independent of system size for both phases. When $q=1$, there is more fluctuation. Although entropic repulsion pushes the average height to order $\sqrt{N}$ above the hard wall at zero height, the variance of the Gaussian free field remains to scale linearly with the total length of the segments, making the EE scale as $\log N$~\cite{PhysRevLett.109.207202}. Note that in a monochromatic version of the embedded 1D chain, all three scalings are sub-area law, which is unique to our construction of filling the 2D lattice with a twisted 1D chain. 

It is worth comparing the EE scalings of the two different 2D models together with a quasi-1D model composed of two layers of Motzkin chains in orthogonal directions that are not interacting with each other, as shown in \cref{fig:snake} (e). Such a construction corresponds to yet another way to define height function in 2D, namely by the Cartesian product of heights along the chains in two different directions. Since the chains in each layer are neither interacting with one another, nor with those in the other direction, the EE scaling in both phase and the at the critical point are simply the corresponding scaling of a single 1D chain multiplied by $L$, as summarized in \cref{tab:table1}. Notice that the result indeed turns out to be drastically different from the spiral construction.

\begin{table}[ht]
	\centering
	\caption{\label{tab:table1}%
		Comparison of the EE scalings (predicted for the spiral-embedded chain) with linear system size $L$ between the two different 2D constructions and the quasi-1D arrays of uncoupled chains.
	}\renewcommand{\arraystretch}{1.5}
	\begin{tabular}{|c |c|c|c |c|}\cline{1-5}
		\textrm{Model}&
		\textrm{Phase}& \textrm{Coupled chains} \cite{ZhaoSixNineteenVertex,Zhang2024quantumlozenge}&
		\textrm{Spiral-embedded chain}& \textrm{Uncoupled chains} \\ \cline{1-5}
		& $q>1$ & $L^2$  & $L^2$ & $L^2$  \\ \cline{2-5}
		Colored & $q=1$  & $L\log L$ & $L$ & $L^{\frac{3}{2}}$ \\ \cline{2-5}
		& $q<1$ & $L$ & $\mathcal{O}(1)$  & $L$\\      \cline{1-5}
		& $q>1$ & $\mathcal{O}(1)$ & $\mathcal{O}(1)$ & $L$  \\ \cline{2-5}
		Colorless & $q=1$ & $L$ \tablefootnote{This is an upper-bound from the zeroth order R\'enyi entropy. The random membrane when bipartitioned gives a Motzkin path on the boundary, giving the dimensionality of the reduced density matrix. The same upper-bound was also obtained for six-vertex models with domain wall boundary conditions using the exact enumeration of alternating sign matrices \cite{Zhang_2023}.} & $\log L$ & $L\log L$   \\ \cline{2-5}
		& $q<1$ & $\mathcal{O}(1)$ & $\mathcal{O}(1)$ & $L$\\
		\cline{1-5}
	\end{tabular}
\end{table}

Unlike the scaling of EE discussed above, which requires somewhat different analyses than the 1D system due to the more complicated bipartition, upper-bounds on the scaling of the spectral gap $\Delta_2(L)$ of the 2D spiral can be straightforwardly deduced from the 1D gap scaling $\Delta_1(L)$ simply by substituting the linear system size with $N=L(L+2)$ giving the relation
\begin{equation}
	\Delta_2(L)= \Delta_1(L^2+2L).\label{eq:con2}
\end{equation}The results are summarized in \cref{tab:table2}. This should be compared with the EE scaling in \cref{tab:table1}. In particular, the $q>1$ phase of the colored model shows that the same EE scaling does not imply the same scaling of spectral gap, and vice versa as shown by the critical point of the colored model.

\begin{table}[ht]
	\centering
	\caption{\label{tab:table2}%
		Comparison of the spectral gap scalings with linear system size $L$ between the two different 2D constructions and the quasi-1D arrays of uncoupled chains, where asterisks denote conjecture scalings. The Greek exponents denote different positive constants.
	}\renewcommand{\arraystretch}{1.5}
	\begin{tabular}{|c |c|c|c |c|}\cline{1-5}
		\textrm{Model}&
		\textrm{Phase}& \textrm{Coupled chains} \cite{ZhaoSixNineteenVertex,mykland2025highlyentangled2dground} &
		\textrm{Spiral chain}& \textrm{Uncoupled chains} \\ \cline{1-5}
		& $q>1$ & $e^{-\alpha L^3}$  & $e^{-\beta L^4}$ & $e^{-\beta L^2}$\\ \cline{2-5}
		Colored & $q=1$  & $L^{-\gamma}$*& $L^{-2\delta}$& $L^{-\delta}$\\ \cline{2-5}
		& $q<1$ &  $\mathcal{O}(1)$*  & $\mathcal{O}(1)$ & $\mathcal{O}(1)$ \\      \cline{1-5}
		& $q>1$ & $e^{-\epsilon L^2}$  & $e^{-\zeta L^2}$  &  $e^{-\zeta L}$\\ \cline{2-5}
		Colorless & $q=1$  & $L^{-\eta}$*  & $L^{-2\theta}$ & $L^{-\theta}$\\ \cline{2-5}
		& $q<1$ & $\mathcal{O}(1)$*  & $\mathcal{O}(1)$ & $\mathcal{O}(1)$ \\
		\cline{1-5}
	\end{tabular}
\end{table}

\section{Tensor network representation of the ground state}\label{sec:TN}

\begin{figure*}[ht]
	\centering
	\includegraphics[width=0.8\linewidth]{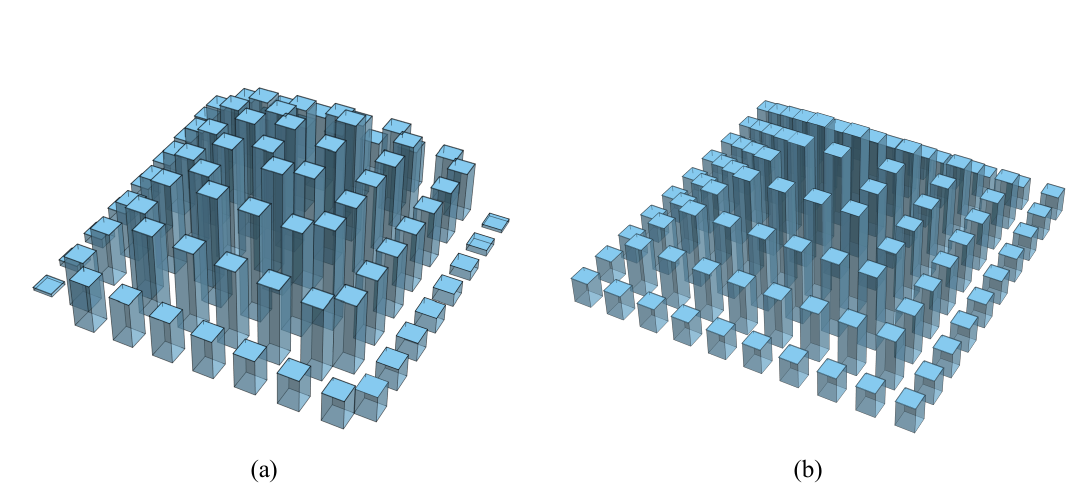}
	\caption{(a) TN representation of the 2D GS of the twisted Motzkin chain, where the contractions in the vertical direction are represented by placing the tensor cubes adjacent to each other, and in the horizontal planes, contraction only happens between neighbors along the spiral. (b) TN representation of the 2D GS of the coupled chain model, where contractions in each horizontal plane happen between neighboring tensor cubes in both directions.}
	\label{fig:TN}
\end{figure*}

The GS of the 1D Motzkin chain has a holographic TN representation \cite{ExactRainbowTensorNetwork}, where each physical spin maps to a column of tensors sequentially contracted with neighbors above and below. The depth of the columns grows linearly from the boundaries of the chain to the center. Twisting the TN from the center results in a 2D TN for the GS of the present model, which has a pyramid shape as shown in \cref{fig:TN} (a). The geometry of the TN is globally the same as the 3D TN for the GS of the coupled Fredkin chain model \cite{mykland2025highlyentangled2dground} shown in \cref{fig:TN} (b). But now the tensors in the bulk have only four internal legs instead of six. The number of bonds crossing the bipartition cut gives the correct scaling of the EE for the $q>1$ phase of the colored model. At the critical point of the colorless model, the TN can be renormalized such that the depth of the columns grow logarithmically instead, giving the correct EE scaling in that case \cite{Alexander2021exactholographic,mykland2025reconcilingtranslationalinvariancehierarchy}.  

The TN representation of the Motzkin and Fredkin GS defined in \cite{ExactRainbowTensorNetwork} actually gives an inverted step pyramid shaped network with the physical external legs at the bottom of each column of tensors. In this section, we define a new TN representation of the Fredkin GS with an upright pyramid shaped geometry, which consist of slightly sparser tensors. This TN is obtained using the tiling approach used in Ref.~\cite{ExactRainbowTensorNetwork}. In this method the correspondence between spin configurations and walks is used to construct the TN representation. This is achieved through an intermediate step where a valid tiling of a set of tiles is associated to each walk present in the GS. In this context, a valid tiling is a tiling where each edge of each tile matches its adjacent tiles. By carefully choosing an appropriate set of tiles, one can assure a one-to-one correspondence between valid tilings and spin configurations present in the GS superposition. For spin configurations present in the GS of the Fredkin model, which are described by Dyck walks, this one-to-one correspondence can be assured by the set of tiles $B_{i}(\bm{c})$ seen in panel (a) of \cref{fig:NewFredkinTN}. In panel (b) and (c) we see that the valid tilings contain arrowed paths indicating the color correlation between spins in the given Dyck walk. A colored spin up (resp.~down) at $i$ is represented by a colored arrowed path entering (resp.~leaving) the tiling at the floor edge of the bottom tile in the tower of tiles at $i$. Clearly, each valid tiling corresponds to a unique way of pairing the degrees of freedom. This set of tiles enables exactly one valid tiling for any given Dyck walk and no valid tiling for non-Dyck walks.
To see this, note that by following the highest arrowed path within each tiling, the corresponding Dyck walk is traced out, see \cref{fig:NewFredkinTN} (b) and (c). It is clear that a valid tiling for a given Dyck walk where this is not the case, is not possible. It is also clear that no valid tiling can be constructed for a non-Dyck walk, as that would necessarily violate arrow continuity at the step where the walk becomes negative (or at the endpoints if the walk starts or terminates at a non-zero height). This establishes the one-to-one correspondence between valid tilings and the spin configurations in the GS. \\

The tensor building up the TN representation is then defined by the set of tiles in the following way: The number of indices of the tensor is determined by the shape of the tiles and each tile represents an index configuration of the tensor giving a nonzero value. Contractions between two tensors in the TN, then correspond to summing over tile configurations matching at that edge. The tiles, valid tilings and the resulting TN, can be seen in \cref{fig:NewFredkinTN}.

\begin{figure*}[hbt!]
	\centering
	\begin{tikzpicture}
		\begin{scope}[shift = {(0, 0.5)}, scale = 0.7]
			
			\drawBone{0, 1.4, 0}{B_{1}(\bm{c})=}{black}{\bm{c}}{}
			\node[scale = 0.85] at (3.1, 1.9) {$=\delta_{\bm{k}_{1}\bm{c}}\delta_{\bm{k}_{2}\bm{0}}\delta_{\bm{k}_{3} \bm{0}}\delta_{\bm{k}_{4}\bm{c}}$};
			\drawBtwo{0, 0, 0}{B_{2}(\bm{c}_{1}, \bm{c}_{2})=}{black}{\bm{c}_{1}}{}{black}{\bm{c}_{2}}
			\node[scale = 0.85] at (3.5, 0.5) {$=\delta_{\bm{k}_{1} \bm{c}_{2}}\delta_{\bm{k}_{2} \bm{c}_{1}}\delta_{\bm{k}_{3} \bm{c}_{1}}\delta_{\bm{k}_{4} \bm{c}_{2}}$};
			\drawBthree{0, -1.4, 0}{B_{3}(\bm{c}_{1}, \bm{c}_{2})=}{black}{\bm{c}_{1}}{}{black}{\bm{c}_{2}}
			\node[scale = 0.85] at (3.7, -0.9) {$=\delta_{\bm{k}_{1} -\bm{c}_{2}}\delta_{\bm{k}_{2} \bm{c}_{2}}\delta_{\bm{k}_{3} -\bm{c}_{1}}\delta_{\bm{k}_{4} \bm{c}_{1}}$};
			\drawBfour{0, -2.8, 0}{B_{4}(\bm{c})=}{black}{\bm{c}}{}
			\node[scale = 0.85] at (3.2, -2.3) {$=\delta_{\bm{k}_{1} -\bm{c}}\delta_{\bm{k}_{2} \bm{c}}\delta_{\bm{k}_{3} \bm{0}}\delta_{\bm{k}_{4} \bm{0}}$};
			\drawBfive{0, -4.2, 0}{B_{5}(\bm{c})=}
			\node[scale = 0.85] at (3.1, -3.7) {$=\delta_{\bm{k}_{1} \bm{0}}\delta_{\bm{k}_{2}\bm{0}}\delta_{\bm{k}_{3}\bm{0}}\delta_{\bm{k}_{4} \bm{0}}$};
			
			%\drawAtwo{2.8, 0, 0}{A_{2}(c)=}{black}{c}{}
			%\drawAthree{5.6, 0, 0}{A_{3}(c)=}{black}{c}{}
			%\drawAfour{0, -1.5, 0}{A_{4}(c)=}{black}{c}{}
			%\drawAfive{2.8, -1.5, 0}{A_{5}(c)=}{black}{c}{}
			
		\end{scope}
		\begin{scope}[shift = {(5, 1.25)}, scale = 0.7]
			% The tiling
			\draw[gray!60, dashed] (0.5, 1.2) -- (0.5, -3.7);
			\draw[gray!60, dashed] (1.5, 1.2) -- (1.5, -3.7);
			\draw[gray!60, dashed] (2.5, 1.2) -- (2.5, -3.7);
			\draw[gray!60, dashed] (3.5, 1.2) -- (3.5, -3.7);
			\drawBone{0, -1, 0}{}{red}{}{}
			\drawBtwo{1, -1, 0}{}{red}{}{}{blue}{}
			\drawBthree{2, -1, 0}{}{red}{}{}{blue}{}
			\drawBfour{3, -1, 0}{}{red}{}{}
			
			\drawBone{1, 0, 0}{}{red}{}{}
			\drawBfour{2, 0, 0}{}{red}{}{}

		\end{scope}
		\begin{scope}[shift = {(5, 1.5)}, scale = 0.7]
			% The Dyck walk
			\draw[thick, red] (0, -4) -- (1, -3);
			\draw[thick, blue] (1, -3) -- (2, -2);
			\draw[thick, blue] (2, -2) -- (3, -3);
			\draw[thick, red] (3, -3) -- (4, -4);
			
			\fill[black] (0, -4) circle (1.5pt);
			\fill[black] (1, -3) circle (1.5pt);
			\fill[black] (2, -2) circle (1.5pt);
			\fill[black] (3, -3) circle (1.5pt);
			\fill[black] (4, -4) circle (1.5pt);
		\end{scope}
		\begin{scope}[shift = {(5, -1.5)}, scale = 0.7]
			% The spin config.
			\draw [thick, red, arrows = {-Stealth[inset=0pt, angle=60 :4pt]}] (0.5, -0.5) -- (0.5, 0);  % Arrow from (1,0) to (0,0)
			\draw [thick, blue, arrows = {-Stealth[inset=0pt, angle=60 :4pt]}] (1.5, -0.5) -- (1.5, 0);  % Arrow from (1,0) to (0,0)
			\draw [thick, blue, arrows = {-Stealth[inset=0pt, angle=60 :4pt]}] (2.5, 0) -- (2.5, -0.5);  % Arrow from (1,0) to (0,0)
			\draw [thick, red, arrows = {-Stealth[inset=0pt, angle=60 :4pt]}] (3.5, 0) -- (3.5, -0.5);  % Arrow from (1,0) to (0,0)
			\node[scale = 0.75] at (0.4, -0.6) {$i = 1$};
			\node[scale = 0.75] at (1.5, -0.6) {$i = 2$};
			\node[scale = 0.75] at (2.5, -0.6) {$i = 3$};
			\node[scale = 0.75] at (3.6, -0.6) {$i = 4$};
			
		\end{scope}
		\begin{scope}[shift = {(9.3, 1.25)}, scale = 0.7]
			% The tiling
			\draw[gray!60, dashed] (0.5, 1.2) -- (0.5, -3.7);
			\draw[gray!60, dashed] (1.5, 1.2) -- (1.5, -3.7);
			\draw[gray!60, dashed] (2.5, 1.2) -- (2.5, -3.7);
			\draw[gray!60, dashed] (3.5, 1.2) -- (3.5, -3.7);
			
			\drawBone{0, -1, 0}{}{red}{}{}
			\drawBfour{1, -1, 0}{}{red}{}{}{blue}{}
			\drawBone{2, -1, 0}{}{blue}{}{}
			\drawBfour{3, -1, 0}{}{blue}{}{}
			
			\drawBfive{1, 0, 0}{}
			\drawBfive{2, 0, 0}{}

		\end{scope}
		\begin{scope}[shift = {(9.3, 1.5)}, scale = 0.7]
			% The Dyck walk
			\draw[thick, red] (0, -4) -- (1, -3);
			\draw[thick, red] (1, -3) -- (2, -4);
			\draw[thick, blue] (2, -4) -- (3, -3);
			\draw[thick, blue] (3, -3) -- (4, -4);
			
			\fill[black] (0, -4) circle (1.5pt);
			\fill[black] (1, -3) circle (1.5pt);
			\fill[black] (2, -4) circle (1.5pt);
			\fill[black] (3, -3) circle (1.5pt);
			\fill[black] (4, -4) circle (1.5pt);
		\end{scope}
		\begin{scope}[shift = {(9.3, -1.6)}, scale = 0.7]
			% The spin config.
			\draw [thick, red, arrows = {-Stealth[inset=0pt, angle=60 :4pt]}] (0.5, -0.5) -- (0.5, 0);  % Arrow from (1,0) to (0,0)
			\draw [thick, red, arrows = {-Stealth[inset=0pt, angle=60 :4pt]}] (1.5, 0) -- (1.5, -0.5);  % Arrow from (1,0) to (0,0)
			\draw [thick, blue, arrows = {-Stealth[inset=0pt, angle=60 :4pt]}] (2.5, -0.5) -- (2.5, 0);  % Arrow from (1,0) to (0,0)
			\draw [thick, blue, arrows = {-Stealth[inset=0pt, angle=60 :4pt]}] (3.5, 0) -- (3.5, -0.5);  % Arrow from (1,0) to (0,0)
			\node[scale = 0.75] at (0.4, -0.6) {$i = 1$};
			\node[scale = 0.75] at (1.5, -0.6) {$i = 2$};
			\node[scale = 0.75] at (2.5, -0.6) {$i = 3$};
			\node[scale = 0.75] at (3.6, -0.6) {$i = 4$};
			
		\end{scope}
		\begin{scope}[shift = {(0, -5.55)}, scale = 0.7]
			
			\node[scale = 0.75] at (-0.9, -0.9) {$i = 1$};
			\node[scale = 0.75] at (0.5, -0.9) {$i = 2$};
			\node[scale = 0.75] at (1.9, -0.9) {$i = 3$};
			\node[scale = 0.75] at (3.3, -0.9) {$i = 4$};
			\node[scale = 0.75] at (5.2, 1.9) {$l = 2$};
			\node[scale = 0.75] at (5.2, 0.5) {$l = 1$};

			\drawB{0, 0, 0}{B(q)}
			\drawB{1.4, 0, 0}{B(q)}
			\drawB{0, 1.4, 0}{B(q)}
			\drawB{1.4, 1.4, 0}{B(q)}
			\drawB{-1.4, 0, 0}{B(q)}
			\drawB{2.8, 0, 0}{B(q)}
			
			\draw (-0.9, -0.35) circle(0.15);
			\draw[thick] (-0.9, -0.5) -- (-0.9, -0.7);
			\draw (0.5, -0.35) circle(0.15);
			\draw[thick] (0.5, -0.5) -- (0.5, -0.7);
			\draw (1.9, -0.35) circle(0.15);
			\draw[thick] (1.9, -0.5) -- (1.9, -0.7);
			\draw (3.3, -0.35) circle(0.15);
			\draw[thick] (3.3, -0.5) -- (3.3, -0.7);

			\drawNoArrowBC{(-0.7, 0.25)}{(-1, 0.25)}
			\drawNoArrowBC{(1.9, 0.25)}{(2.2, 0.25)}
			\drawNoArrowBC{(0, 0.95)}{(-0.3, 0.95)}
			\drawNoArrowBC{(1.2, 0.95)}{(1.5, 0.95)}
			
			\drawNoArrowBC{(-0.45, 0.5)}{(-0.45, 0.8)}
			\drawNoArrowBC{(0.25, 1.2)}{(0.25, 1.5)}
			\drawNoArrowBC{(0.95, 1.2)}{(0.95, 1.5)}
			\drawNoArrowBC{(1.65, 0.5)}{(1.65, 0.8)}

		\end{scope}
		\begin{scope}[shift = {(6.8, -5.4)}, scale = 0.7]
			\drawB{-0.3, 0, 0}{B(q)}
			\node[scale = 0.85] at (0.25, -0.5) {$\bm{k}_{1}$};
			\node[scale = 0.85] at (-0.75, 0.5) {$\bm{k}_{2}$};
			\node[scale = 0.85] at (0.25, 1.5) {$\bm{k}_{3}$};
			\node[scale = 0.85] at (1.25, 0.5) {$\bm{k}_{4}$};
			
			\drawNoArrowBC{(1.2, 0.15)}{(1.2, 0.55)}
			\node[scale = 0.85] at (3, 0.6) {$ = $};
			\node[scale = 0.85] at (2.4, 0.1) {$ \bm{k} $};
			\node[scale = 0.85] at (3.6, 0.6) {$ \delta_{\bm{k}\bm{0}} $};
			
			\draw[thick] (4.7, 0.65) -- (4.7, 0.9);
			\draw (4.7, 0.5) circle(0.15);
			\draw[thick] (4.7, 0.35) -- (4.7, 0.1);
			\node[scale = 0.85] at (6.2, 0.6) {$ =\hat{1} - |\bm{0}\rangle\langle\bm{0}| $};

		\end{scope}
		\begin{scope}[shift = {(1.5, -3)}]
			\node at (0, 0) {(a)};
			\node at (5, 0) {(b)};
			\node at (9.25, 0) {(c)};
			\node at (-0.65, -3.9) {(d)};
			\node at (7.25, -3.9) {(e)};
			
		\end{scope}
	\end{tikzpicture}
	\caption{(a) The five different tiles $B_{i}(\bm{c})$ for the new Fredkin tensor network. The tiles are also defined as rank-4 tensors in terms of Kronecker deltas, where the indices $\bm{k}_{i}$ are defined as for $B(q)$ in panel (d). We have $\bm{c} = (1, 0)$ for red arrow, $\bm{c} = (0, 1)$ for blue arrow. No arrow corresponds to $\bm{0}$. (b)-(c) Valid tilings corresponding to the maximal and minimal height Dyck walk for the $L = 4$ system. (d) TN representation of the GS of the single Fredkin chain for $L = 4$. $l$ denotes the different levels of the holographic TN. (e) The constituent tensors of the TN, where the white circle projects out the index value $\bm{0}$ corresponding to no arrow in the tilings.}
	\label{fig:NewFredkinTN}
\end{figure*}

The nonzero entries of the tensor in terms of the deformation parameter $q$ are determined by counting the number of tiles in a tiling and considering the area under the corresponding Dyck walk. By weighing each tile with an appropriate value of the deformation parameter $q$, one can ensure that contracting the TN yields the weighted superposition of spin configurations in the GS. This gives a rank-4 tensor $B(q)$ as
\begin{equation}
	\begin{split}
		B(q) = \sum_{\bm{c}_{i} = (1, 0), (0, 1)}&[\sqrt{q}B_{1}(\bm{c}_{1}) +qB_{2}(\bm{c}_{1}, \bm{c}_{2})
		+ q B_{3}(\bm{c}_{1}, \bm{c}_{2}) ,+\sqrt{q}B_{4}(\bm{c}_{1})+B_{5}(\bm{c}_{1}) ].
	\end{split}
	\label{eq: NewFredkinSpinChainTensor}
\end{equation}
The indices $\bm{k}_{1},\cdots, \bm{k}_{4}$ of $B(q)$ are defined in panel (d) of \cref{fig:NewFredkinTN} and have been suppressed in \cref{eq: NewFredkinSpinChainTensor}. We use a vector notation for the indices to keep track of the color of the arrows, which is a notation conveniently extended to the tensors in the TN for the 2D systems \cite{mykland2025highlyentangled2dground}. It is clear that one can simply replace the different vectors with a unique number, to arrive at the more familiar way of denoting index configurations by numbers. Note also that we use a minus sign to indicate arrows leaving (resp. coming in) at the $\bm{k}_{1}$ (resp. $\bm{k}_{3}$) index. In \cref{eq: NewFredkinSpinChainTensor}, we consider two colors, namely red, $\bm{c} = (1, 0)$, and blue, $\bm{c} = (0, 1)$. Note that each of the tiles $B_{i}(\bm{c})$ can be defined as a rank-4 tensor expressed as a product of Kronecker deltas, as seen in panel (a) in \cref{fig:NewFredkinTN}.  The shape of the resulting TN naturally becomes that of the valid tilings, giving an upright step pyramid shape as seen in panel (d) in \cref{fig:NewFredkinTN}. Note that the boundary tensors $\delta_{\bm{k}, \bm{0}}$ ensures that ``no arrows flow out'' of the TN, except at the bottom edges. Due to the tile without arrows, we must project out the index value corresponding to an empty edge at the bottom tile. We must therefore include projectors in the TN, as seen in \cref{fig:NewFredkinTN} (d). This is similar to the rainbow TN presented with projectors in Ref. \cite{ExactRainbowTensorNetwork}. \\

Finally, we mention that one can restrict the tensors at even and odd sites by taking into account that a Dyck walk can only reach an even (resp. odd) height after an even (resp. odd) number of steps. This implies that in valid tilings the tile above an even (resp. odd) position $i$, at an odd (resp. even) level $l$ can never be the $B_{1}(\bm{c})$ tile, as that would correspond to a walk reaching an even height (resp. odd) after an odd (resp. even) number of steps. Due to the same reason the $B_{4}(\bm{c})$ tile can never be placed above an even (resp. odd) position, at an even (resp. odd) height. This means that one can build a TN where the tensors are built up of 4 tiles, instead of 5, at the expense of having two different tensors. One is $B(q)$ without $B_{1}(\bm{c})$ and the other one is $B(q)$ without $B_{4}(\bm{c})$. This gives a TN with slightly sparser tensor than the one presented in Ref. \cite{ExactRainbowTensorNetwork}, at the cost of having two different tensors in the TN.

\section{Junction extension and scaling conjecture}\label{sec:junction}

\begin{figure*}[ht]
	\centering
	\includegraphics[width=0.6\linewidth]{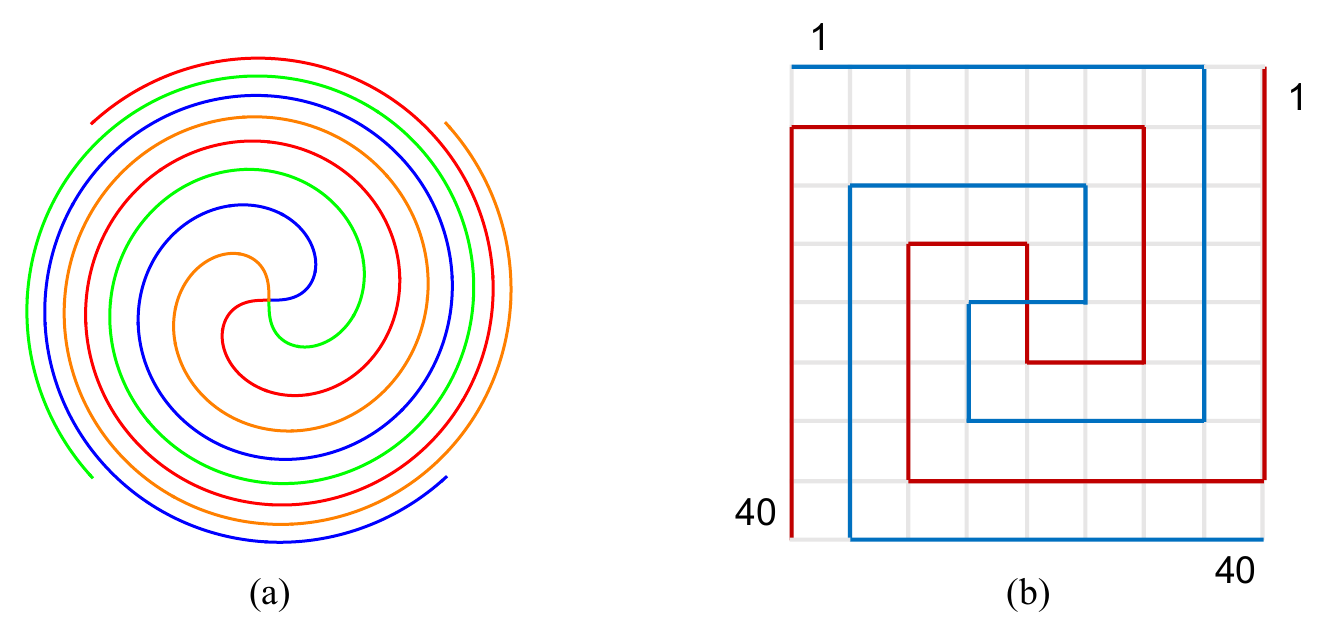}
	\caption{(a) Two Fermat-spirals sharing the same origin, with each of their two branches depicted in a different color. (b) The corresponding Hamiltonian path on a square lattice, which can be considered a junction of two spiral Motzkin chains at the center.}
	\label{fig:2spiral}
\end{figure*}

The spiral model has two distinct geometrical structures. The occupied degrees of freedom fill a two-dimensional lattice region, so their number scales as $L2$. In contrast, the active interaction graph of the single-spiral model is a path and therefore has one-dimensional graph connectivity. The construction should consequently be viewed as a one-dimensional local Hamiltonian equipped with a two-dimensional spatial embedding and a family of ambient-space bipartitions. Its purpose is not to construct a fully two-dimensional interaction network, but to isolate how embedding and subsystem geometry alter the entanglement observed across spatial cuts.

\begin{figure*}[ht]
	\centering
	\includegraphics[width=\linewidth]{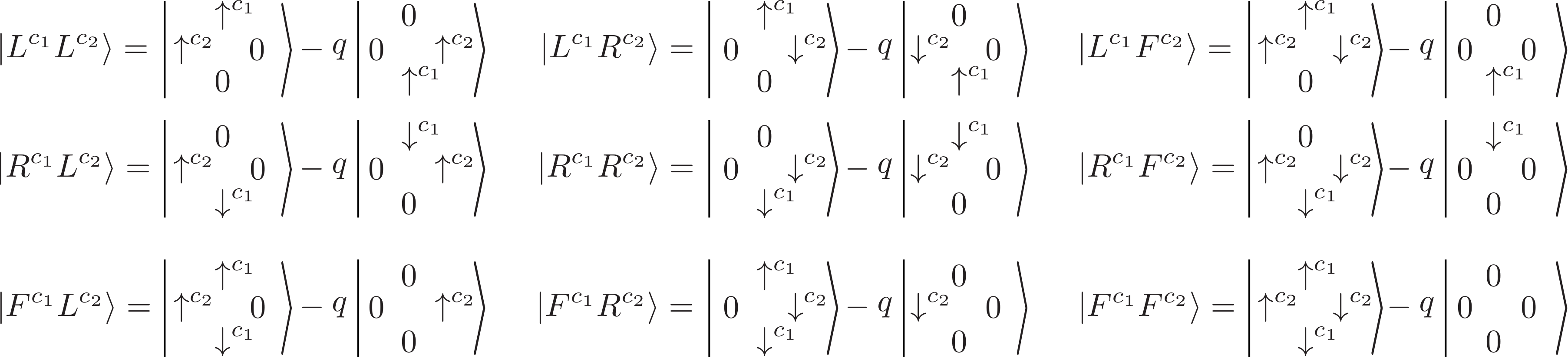}
	\caption{Projection operators acting on the four neighboring spins at the center of the lattice are defined in terms of these vectors.}
	\label{fig:center}
\end{figure*}

The construction can be modified to include a junction of two Motzkin chains, as shown in \cref{fig:2spiral} (b). The two constituent chains are depicted in different colors. The junction introduces a local branching point but still does not produce two-dimensional graph-volume growth. Along each chain, the spins interact via the same Motzkin Hamiltonian as before, except for the four central spins, where the local Hamiltonian is replaced by
\begin{equation}
	H_{N,N}=\sum_{M_1,M_2=R,L,F}\sum_{c_1,c_2=1}^s\frac{1}{1+q^2}|M_1^{c_1}M_2^{c_2}\rangle\langle M_1^{c_1}M_2^{c_2}|,
\end{equation}defined in terms of the vectors illustrated in \cref{fig:center}. This local interaction is exactly the same as in the coupled 19-vertex construction of Ref.~\cite{ZhaoSixNineteenVertex}. Here, however, in contrast to a fully translationally invariant implementation of such a term, the more intricate coupling is applied only once, at the spiral junction. As an interesting future direction, one could explore interpolations between the Motzkin spiral model and the 19-vertex coupled-chain model by allowing such junctions to proliferate throughout the lattice.

Operationally, this interaction simultaneously raises or lowers the height between the $N$th and $(N+1)$st spins on both spirals, whenever this move is allowed for both chains. The ground state therefore remains a frustration-free superposition of all Motzkin configurations on the two chains, subject to the additional constraint that their midpoints share the same height. Since the leading contribution to the EE comes from Bell-type pairing in the color degrees of freedom, which remain independent between the chains, the EE scaling is still governed by the average height scaling along each chain. There is no reason to expect the scaling in the $q>1$ or $q<1$ phases to change qualitatively. The coupling at the junction does reduce height fluctuations on both chains, and not only at the center; however, the constraint only enforces height matching between two 1D Motzkin walks. This is very different from the coupled-chain models of Refs.~\cite{ZhaoSixNineteenVertex,Zhang2024quantumlozenge}, where the walks are promoted to effectively 2D random surfaces, leading to a much stronger suppression of fluctuations \cite{DeuschelZeitouni,c6e74874-6cdf-3049-8d6c-6fbee4e7dc77}. Consequently, we do not expect the EE scaling to be modified. 

A more rigorous estimate of this scaling can be made based on the assumptions leading to \eqref{eq:Shrelation}, which relates the leading EE contribution to the average height at the junction $n$. Using the scaling limit of Motzkin paths derived in the Supplementary Material of Ref.~\cite{PhysRevLett.109.207202}, we have that the probability density of $n$ behaves as $p^\mathrm{C}_n\sim n^2e^{-\frac{3n^2}{N}}$, for a single Motzkin chain after $N$ spins. This gives the joint probability density for height at the junction of two chains $p^\mathrm{J}_n\sim n^4e^{-\frac{6n^2}{N}}$. The maximum of the joint probability density is still achieved at $n^*= \sqrt{N/3}$ as is the case of $p^\mathrm{C}_n$ for a single Motzkin chain. So using the same saddle-point approximation as before to compute the expectation value of the height at the junction, we get the same EE scaling as the Motzkin spiral model with $S(L)\sim \sqrt{N}\sim L$ for $q=1$, as $4N=L(L+2)$.

%
%%
%%%
\section{Conclusion}
\label{sec:Concl}
%%%
%%
%
We have studied the entanglement properties of 1D Motzkin chains embedded in a 2D square-lattice. In particular, these embeddings exhibit the same volume-scaling of bipartite EE for the $q>1$ phase, yet different scalings for the critical point and $q<1$ phase from those found in earlier constructions based on coupled chains with genuine 2D interactions. Our results thus provide a complementary route to obtaining nontrivial EE scaling for subsystems defined on a 2D lattice, without requiring a fully coupled 2D interaction network.

Our constructions are based on simple 1D Motzkin Hamiltonians embedded into the 2D lattice via (i) a single Fermat-spiral-like Hamiltonian path, and (ii) two Motzkin chains arranged in spiral layouts with a coupled junction at the center. The active bonds follow the spiral paths, with their orientations varying across the four quadrants of the square lattice. Near the diagonals, chain bonds can be oriented along different lattice directions, reflecting the geometry of the embedding. These constructions illustrate how the spatial arrangement of a 1D chain can give rise to distinctive entanglement scaling for bipartitions of a 2D lattice.

Besides involving much simpler interactions than fully coupled 2D construction, another key advantage of these constructions is that they allow us to infer the EE and spectral-gap scaling directly from known 1D results via the conversion relations \eqref{eq:con1} and \eqref{eq:con2}, which constitute the main technical results of this work. While the spectral gap relation is exact, the entanglement relation is subtler because it relies on the assumption of approximately monotonic height variation within each segment, an approximation supported by both analytical arguments and numerics in Sec.~\ref{sec:EE}. The same approximation predicts that coupling two Motzkin spirals at a central junction does not change the scaling of EE. Regardless of the details of these approximations, the embedded chain inherits the one-dimensional ground-state transition as $q$ is varied. The scaling of its ambient-cut entropy is predicted to change accordingly. In particular, for the spiral model even the area-law phase is bounded by a constant, further motivating the exploration of how different embeddings of 1D models affect EE scaling on a 2D lattice.

There are many additional properties of the spiral constructions that remain to be understood. In this work we have emphasized EE and spectral-gap scaling, leaving a systematic study of correlation functions for future investigation. Notably, despite the strictly anisotropic interactions within each quadrant, the embedding produces a staggered one-point spin profile in the radial direction, where there is no direct interaction; this phenomenon is discussed in \cref{sec:corr}. For the Motzkin junction model, one can also probe spin correlations between different spirals, in close analogy with studies of quantum wire junctions \cite{wirejunction,CRAMPE2013526,Juhász_2018,fddq-8lzl}. Although our explicit examples are based on Motzkin and Fredkin chains, the spiral embedding should apply equally well to other highly entangled one-dimensional systems, such as rainbow chains \cite{Vitagliano_2010,Ramírez_2014,Ramírez_2015}. It would be particularly interesting to test whether the same conversion formulas for EE can produce new classes of 2D scaling behaviors. For instance, Ref.~\cite{ZhaoSixNineteenVertex} proposed generic power-law EE scaling by introducing an inhomogeneous deformation parameter $q$, although no local parent Hamiltonian was found for that ground state.

Finally, there are natural limitations to how far the spiral approach can be extended. In our 2D layouts, volume-scaling EE arises because the color degrees of freedom preferentially form Bell pairs with sites related by point reflection about the system’s center. It is not obvious how to generalize this mechanism to three dimensions, where one would need an inversion-symmetric path that visits all lattice sites in three-dimensional space. By contrast, coupled-chain constructions based on tiling can, in principle, be extended to arbitrary dimensions. Together, these models of area-law–violating GSs enrich our understanding of the interplay between EE scaling and spectral gaps, and of the tradeoff between entanglement as a computational resource and robustness against noise. A long-term goal in this direction is to formulate a generalization of the area law to gapless systems, relating the decay of the spectral gap to the growth of EE with system size.

%
% Each of the commands below will create an unnumbered section with the appropriate heading.
% Remove any sections that are not relevant for your article.
% All sections except suppdata will be removed if the [anonymous] option is used.
% See iopjournal-guidelines.pdf for more information.
%

\ack{ZZ thanks Shankar Balasubramanian, Soonwan Choi, Israel Klich, Ethan Lake, and Germ\'an Sierra for valuable discussions.}

%\funding{Sample text inserted for demonstration.}
% This section is a list of funder names and grant numbers

%\roles{Sample text inserted for demonstration.}
% List author names and the contributions made to the article, using terms from the NISO Contributor Roles Taxonomy (CRediT) https://credit.niso.org

%\data{Sample text inserted for demonstration.}
% For more information on IOP Publishing's research data policy see: https://publishingsupport.iopscience.iop.org/questions/research-data/

%\suppdata{Sample text inserted for demonstration.}
\appendix

\section{Typical height change in a segment of Motzkin path} \label{sec:reflection}

\begin{figure*}[hbt!]
	\centering
	\includegraphics[width=0.5\linewidth]{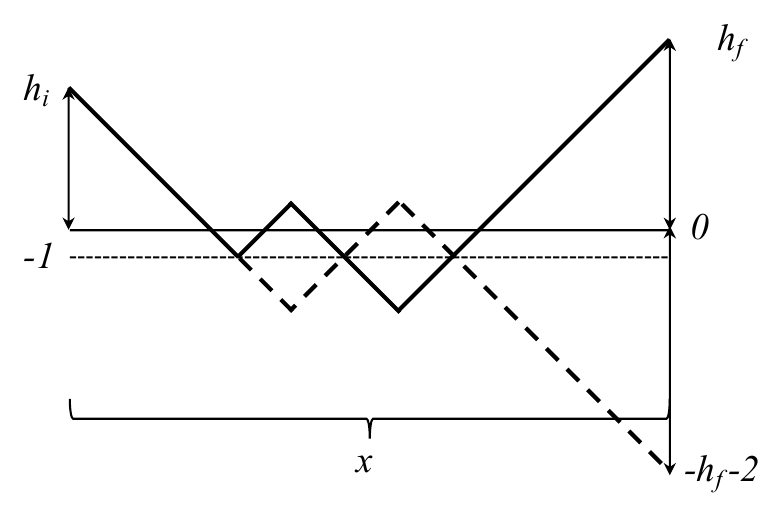}
	\caption{The number of segments of Motzkin paths that start at height $h_i$ and end at height $h_{f}$ over $x$ steps without having negative heights in the middle can be enumerated with Andr\'e's reflection method.}
	\label{fig:reflection}
\end{figure*}

In this appendix, we determine the typical height change in a segment of Motzkin path by finding the most probable height change using combinatorics. The number of Motzkin path segments that start at height $h_{i}$ and end at height $h_f$ over a horizontal distance of $x$ can be enumerated using Andr\'e's reflection method, see \cref{fig:reflection}. Such paths meeting the Motzkin condition of non-negative height everywhere is equal to the total number of random walks subtracted by those that does reach height $-1$ somewhere. The latter are in one-to-one mapping with paths that are mirror images about the horizontal line $h=-1$ after the first time the path reaches height $-1$. These paths are in turn enumerated by the random walks from height $h_i$ to height $-h_f-2$.

For simplicity, we first count such Dyck paths, which are Motzkin paths without the option of flat moves, as
\begin{equation}
	\sqrt{p_\mathrm{D}(h_f,h_i,x)}= \binom{x}{\frac{x+h_f-h_i}{2}}-\binom{x}{\frac{x+h_f+h_i}{2}+1}\approx 
	\sqrt{\frac{2}{\pi x}}2^{x}\left(e^{-\frac{(h_f-h_i)^2}{x}}-e^{-\frac{(h_f+h_i)^2}{x}}\right)
	\label{eq:prob_dist}
\end{equation}where the limit $x\gg h_i,h_f\gg 1$ has been used. The most probable height change $h_f^*-h_i$ is determined by $p_\mathrm{D}'(h_f^*,h_i,x)=0$, which is solved from
\begin{equation}
	e^{\frac{4h_ih_f^*}{x}}=\frac{h_f^*+h_i}{h_f^*-h_i}.
\end{equation}
Motzkin paths starting and ending at finite height can be counted analogously, resulting in
\begin{equation}
	\sqrt{p_\mathrm{M}(h_f,h_i,x)}\propto e^{-\frac{3(h_f-h_i)^2}{2x}}-e^{-\frac{3(h_f+h_i)^2}{2x}},
\end{equation}and the typical height change determined by
\begin{equation}
	e^{\frac{6h_ih_f^*}{x}}=\frac{h_f^*+h_i}{h_f^*-h_i}.
\end{equation}
The leading contribution to EE is proportional to the total net height change between the two subsystems
\begin{equation}
	S\sim\sum_{i=0}^{\frac{L}{2}-1}|\langle h_{2i+1}-h_{2i}\rangle|\propto\int\mathcal{D}[h]\left[\sum_{k=1}^{\frac{L}{2}}(-1)^{k}h_{k-1}+\sum_{k=\frac{L}{2}}^{L-1}(-1)^{k}h_{k}\right],
	\label{eq:pathintegral}
\end{equation}where $h_i$ are the height variable at the $L$ endpoints of the segments resulting from the bipartition on the 2D lattice. In the second form of the equation the contribution from the segments lying in the first half of the chain and contributions from segments lying in the second half of the chain have been split into two sums, using that in the first half of the chain the height increases, while in the second half of the chain it decreases. Throughout the paper, $L$ assumed an integer multiple of 4 without loss of generality. The functional integral over the height configuration $h(x)$ is done by integrating over the height variable at discrete locations weighted by corresponding probabilities
\begin{equation}
	\mathcal{D}[h]=p(h_1)\prod_{i=1}^{\frac{L}{2}-1}dh_ip(h_{i+1},h_{i},|4i-2L|)\prod_{j=\frac{L}{2}}^{L-2}dh_jp(h_{j+1},h_{j},|4(j+1)-2L|)p(h_{L-1}),
\end{equation}with the probability distribution of boundary segments are given by \cite{PhysRevLett.109.207202,FredkinSpinChain}
\begin{equation}
	p(h)=\begin{cases}
		\begin{aligned}
		\frac{2^{4L+3}h^2}{\pi x}e^{-\frac{h^2}{L}}, &\quad  \text{ for Dyck paths};\\
		\frac{2^{4L+3}h^2}{\pi x}e^{-\frac{3h^2}{2L}}, &\quad  \text{ for Motzkin paths}.\\
		\end{aligned}
	\end{cases}
\end{equation}

\section{Emergent staggered spin profile}
\label{sec:corr}

\begin{figure*}
	\centering
	\includegraphics[width=0.8\linewidth]{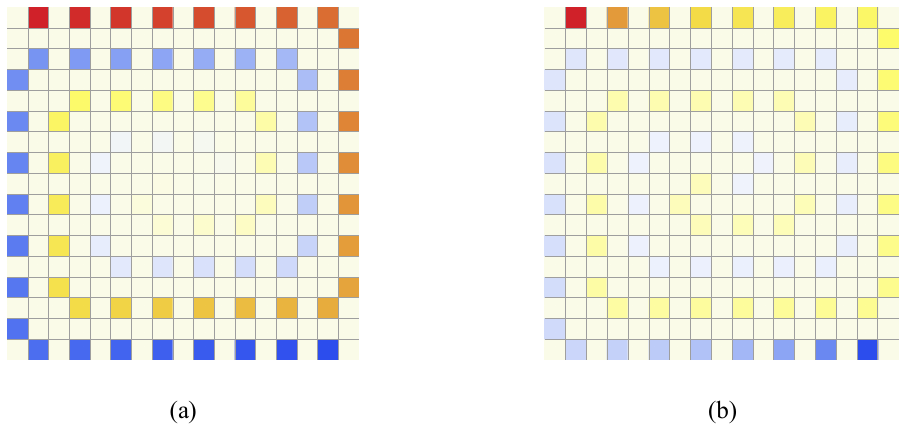}
	\caption{Order parameter $\langle S^z\rangle$ for (a) $q > 1$, and (b) $q=1$. Warm color and cold colors denote respective spin up and down, and darker colors correspond to larger magnitude. The neutral muted shade represents 2D lattice sites not occupied by degrees of freedom of the 1D chain.}
	\label{fig:order}
\end{figure*}

We show in this appendix that the spiral layout of the Motzkin on the 2D lattice induces an antiferromagnetic order of the spin degree of freedom within each quadrant along the direction that there is no interaction. The expectation value of the spin in a 1D chain has been computed for $q=1$ from combinatorial methods \cite{10.1063/1.4977829,Menon:2024vic}, field theoretic approach in the scaling limit \cite{mykland2025highlyentangled2dground}, and the transfer matrix method \cite{mykland2025reconcilingtranslationalinvariancehierarchy}. All three approaches lead to the same scaling behavior as $r^{-1/2}$, where $r$ is the distance from the closer boundary, with up spin on the first half, and down spin on the second. When twisted and embedded on the 2D lattice, an induced antiferromagnetic order appears with modulated amplitude that decays from the boundary to the center, as shown in \cref{fig:order} (b).

As was shown in Ref.~\cite{mykland2025reconcilingtranslationalinvariancehierarchy}, due to the opposite boundary conditions, the $q>1$ ordered phase forms a domain wall between the two regions with positive and negative values of the order parameter. The correlation length in this case is reflected in the domain wall thickness $\xi$ by
\begin{equation}
	\langle S^z_k\rangle\propto \tanh{\frac{N-k}{\xi}},
\end{equation}which can be derived from the Ginzburg-Landau theory \cite{mykland2025reconcilingtranslationalinvariancehierarchy}. The length scale depends on the deformation parameter as $\xi\propto 1/\log q$, which was obtained from numerical results by a matrix product state calculation \cite{mykland2025reconcilingtranslationalinvariancehierarchy}. \cref{fig:order} (a) depicts the twisting induced antiferromagnetic order in the direction orthogonal to 1D chain within each quadrant of the 2D lattice for $\xi=2N$.
%\section*{References}
\bibliographystyle{iopart-num}
\bibliography{snake}

\end{document}